\documentclass[prb,superscriptaddress, aps, showpacs, amsmath, amssymb, nofootinbib,twocolumn]{revtex4-1}

\usepackage{graphicx}
\usepackage{epstopdf}
\usepackage{dcolumn}
\usepackage{bm}
\usepackage{bbm}
\usepackage{amssymb}
\usepackage{amsfonts}
\usepackage{amsthm}
\usepackage{dsfont}
\usepackage{amsmath}

\newcommand{\be}{\begin{equation}}
\newcommand{\ee}{\end{equation}}
\def\ba{\begin{array}}
\def\ea{\end{array}}

\begin{document}
\title{Multipoint entanglement 
in disordered systems}
\author{Javier M. Mag\'an}
\affiliation{Institute for Theoretical Physics \emph{and} Center for Extreme Matter and Emergent Phenomena, Utrecht University, 3508 TD Utrecht, The Netherlands }
\author{Simone Paganelli}
\affiliation{Dipartimento di Scienze Fisiche e Chimiche, Universit\`{a} dell'Aquila, via Vetoio, I-67010 Coppito-L'Aquila, Italy}
\affiliation{International Institute of Physics, Universidade Federal do Rio Grande do Norte, 59012-970 Natal, Brazil}
\author{Vadim Oganesyan}
\affiliation{Department of Engineering Science and Physics,
College of Staten Island, CUNY, Staten Island, NY 10314, USA}
\affiliation{Physics program and Initiative for the Theoretical Sciences, The Graduate Center, CUNY, New York, NY 10016, USA}

\date{\today}

\begin{abstract}
We develop an approach to characterize excited states of 
disordered many-body systems using spatially resolved structures of entanglement.
We show that the behavior of the mutual information (MI) between two parties of a many-body system can signal a qualitative difference between thermal and localized phases -- MI is finite in insulators while it approaches zero in the thermodynamic limit in the ergodic phase. Related quantities, such as the  recently introduced Codification Volume (CV), are shown to be suitable to quantify the correlation length of the system. These ideas are illustrated using prototypical non-interacting wavefunctions of localized and extended particles and then applied to characterize states of strongly excited interacting spin chains.
We especially focus on evolution of spatial structure of quantum information between high temperature diffusive and many-body localized phases believed to exist in these models. We study MI as a function of disorder strength both averaged over the eigenstates and in time-evolved product states drawn from continuously deformed family of initial states realizable experimentally. As expected, spectral and time-evolved averages coincide inside the ergodic phase and differ significantly outside.
We also highlight dispersion among the initial states \emph{within} the localized phase -- some of these show considerable generation and delocalization of quantum information. 
\end{abstract}


\maketitle

\section{Introduction}
Transmission of classical information is inextricably tied to transport of conserved or nearly conserved observables, such as magnetization or energy. Information channels, esp. their signal-to-noise properties, are formally characterized using classical information theory notions, e.g. Shannon capacity. In practice, communications are best accomplished through media supporting nearly ballistic propagation, often sufficiently described by very few parameters, e.g. pulse dispersion and attenuation (see Refs. \onlinecite{Bose2007,Apollaro2015,Lorenzo2015} and references therein).
By contrast, quantum information is encoded
in 
the wavefunction of the many-body system (e.g. its entanglement), where it is often subject to "dephasing", i.e. degradation by so called $T_2$ processes that ordinarily take place with no exchange of energy \cite{}. Transport of entanglement can be divorced from transport of classical information.

T
hree common paradigms of transport -- ballistic transport, diffusion and localization -- may be clearly identified from time dependence of observables, e.g. spread of wavepackets. 
By contrast, temporal evolution of quantum information has not been thoroughly classified. In recent years, closely related properties, such as von Neumann and Renyi bipartite entanglement entropies of subsystems of isolated many-body systems have received serious attention, especially regarding its  area-law behavior in the ground state \cite{Srednicki1993,Amico2008,Eisert2010,Calabrese2004,Callan1994} and its possible deviations \cite{Wolf2006,Ramirez2014,Gori2015}.  One prominent result is the (Lieb-Robinson) linear in-time growth of entanglement in ballistic and diffusive systems \cite{lieb}. By contrast, there exist localized interacting many-body systems, known as "many-body localized" phases  \cite{dalessio2016,basko,vadim3, znidaric2008,monthus2010, Berkelbach2010,  huse2, buccheri2011, canovi2011, deluca2013} , whose subsystems' entanglement grows only logarithmically in time\cite{moore, dmitry, huse,dechiara2006, znidaric2008}.

When the many-body system is sufficiently excited (e.g. to finite statistical entropy density vis-a-vis finite excitation energy density) these behaviors are encoded in the entanglement of many-body eigenstates -- exhibiting surface law in localized phases and volume law otherwise. Latter behavior is usually related to "eigenstate thermalization hypothesis"\cite{vadim3,srednicki, deutsch,rigol2008,Magan2015}. On the contrary, MBL phases display a - possibly partially- localized spectrum and  a correspondent mobility edge\cite{Baygan2015,Bera2015,Mondragon2015,Xu2015,Naldesi2016} . The need for a full knowledge of the entire spectrum  makes the study of the MBL phases numerically extremely challenging \cite{Pollmann2016,Luitz2015} . While these are predominant ``paradigms'' of transport and entanglement growth more ornate possibilities have been discussed, e.g. non-universal powerlaws from bottlenecks in low dimensional systems (see, e.g. ref. \onlinecite{Gopalakrishnan2015a} and refs. therein), entanglement on expander graphs \cite{us3,avi1,avi2}, or entanglement evolution in fully non-local systems \cite{magan2,magan3}.

In this work we focus on developing tools for characterizing spatial (and temporal) patterns of entanglement, especially at short distances, thus complementing earlier and ongoing works
\cite{Grover2014,Pollmann2013,Xu2015,Singh2016,Campbell2016,Iemini2016,Detommasi2016,Goold2015,Goold2016,Khemani2016,Zhang2016}, where different aspects of correlations/entanglement were argued to undergo a phase transition concomitant with many-body localization-delocalization transition\cite{Chandran2015}.     There are two basic physical motivations here: (i) properly analysed and sampled short range correlations are expected (based on ETH) to be as sharp observables as long range entanglement structures;
(ii) few site entanglement entropy and quantum information is measurable (inferable, to be precise) in principle in experiments, from complete tomography of the reduced density matrix of small subsystems.
In the longer term we aim to develop a comprehensive quantum hydrodynamic description of excited matter, one that encapsulates the description of transmission of classical information (ordinary transport) and its quantum counterpart.

This paper is organized as follows: definitions of multipoint entanglement measures are made in the next Section, these definitions are applied to localized and extended single particle states in the following Section III, many-body behaviors are analysed in Section IV, including dispersion of quantum information with respect to initial conditions in quantum quenches. We close with a discussion of possible broader implications of our results and some open questions.
\section{Definitions}
We consider dividing the ``Universe'' into three mutually non-overlapping subsystems of physical degrees of freedom, A, B and C, e.g. segments of a spin-chain. To each  of these we may associate an Hermitean operator algebra closed under multiplication
\begin{align}
\rm{Universe}&\to \mathcal{U}
\\
\rm{A}&\to \mathcal{A}
\\
\rm{B}&\to \mathcal{B}
\\
\rm{C}&\to \mathcal{C}
\\
\mathcal{A}\cup\mathcal{B}\cup\mathcal{C}&= \mathcal{U}
\\
\mathcal{A}\cap\mathcal{B}=\mathcal{A}\cap\mathcal{C}&=\mathcal{B}\cap\mathcal{C}=\mathcal{I},
\end{align}
where $\mathcal{I}$ is the (global) identity.
For example if A and B corresponds to sites 13 and 17 in the spin 1/2 chain of length 21, there are 4 operators in each of $\mathcal{A}$ and $\mathcal{B}$ (the Pauli matrices) and  $4^{19}$ operators in $\mathcal{C}$ of the form $\sigma^x_1 \otimes \mathds{1}_2 \otimes \sigma^z_3\cdots \sigma^z_{21}$.  More generally we will have $d$ dimensional physical degrees of freedom (by this definition $d=4$ for S=1/2). These operator algebras form the natural orthonormal basis for decomposing the corresponding reduced density matrices for A,B,C as they encode the values of various \emph{measureable} multispin correlations from which one may in principle (and in experiment) reconstruct the density matrix and compute its information theoretic content.

Treating subsystems A and B as two putative communicating parties and subsystem C as the ``environment'' we may ask to quantify the amount of quantum information shared by A and B in a pure state (or density matrix) by computing the distance between reduced density matrix of A and B together and the product of reduced density matrices of them individually, $||\rho_{AB}-\rho_{A}\otimes \rho_{B}||$.
One good notion of this quantum distance is the quantum relative entropy 
\begin{equation}\label{relativedefinition}
S(\rho\Vert \sigma)=\textrm{Tr}\rho (\log\rho-\log\sigma)\;,
\end{equation}
because it bounds other common definitions \cite{eisert,cirac}. Since in our case $A$ and $B$ are disjoint, the relative entropy coincides with the mutual information $I(A,B)=S_A+S_B-S_{AB}$ between A and B, defined using (von Neuman) entropy of the reduced density matrices
\begin{equation}\label{relativeI}
S(\rho_{AB}\Vert \rho_{A}\otimes 
\rho_{B})= S_E(\rho_{A})+S_E(\rho_{B})-S_E(\rho_{AB})= I(A,B).
\end{equation}
The MI gives the total amount of quantum correlations between two subsystems \cite{groisman2005} and it is a measure of how much we can learn of $A$ by studying $B$ and vice versa.  Presence of MI relies on entanglement but may also be suppressed by excessive entanglement. As such we expect it to be particularly sensitive to changes between the patterns of entanglement e.g. from area to volume law.

Suppose we choose subsystem A and ask to \emph{maximize} the amount of mutual information over all possible B's.  Given strong subadditivity:
\be
I(A,B\cup C)\geq I(A,B)\;,
\ee
the solution is clearly $\bar{A}$, the complement of A.  Now, suppose we introduce a tolerance parameter $\epsilon$ and ask to find the smallest subsystem $B_{\epsilon}$ s.t. 
\be
I_{\rm max}-I(A,B_{\epsilon})<\epsilon.
\ee
The number of spins (or degree of freedom) of this minimum $B_{\epsilon}$ defines a \emph{codification volume}. No subsystem smaller than $B_{c}$ contains the required information about A. For globally pure states, we may attach a physical meaning to $\epsilon$ by noting that the \emph{deficit} information 
\be\label{eq:ic}
I(A,C)=I(A,\bar{A})-I(A,B_{\epsilon})<\epsilon
\ee
is encoded in the subsystem $C$ external to $A$ and $B$.  Thus, $\epsilon$ is an upper bound on information not captured by the codification volume $B_{\epsilon}$.  As explained in Ref. \onlinecite{uscod}, the size this minimum $B_{\epsilon}$ is a good measure of information localization. As such it is expected to diverge  from localized to delocalized phases. For many highly entangled states of matter, especially ones displaying extensive (volume-law) entanglement, the notion of codification volume may not have any practical significance since it diverges in the thermodynamic limit due to large information deficit. On the other hand we will explore some nontrivial examples for which exponential localization of information is realized, e.g. 
\be\label{eqn:expodecay}
I(A,C) \propto e^{-L_\beta/\xi},
\ee
where $L_\beta$ is the linear size of $\beta$ and $\xi$ is the characteristic length of localization of quantum information (which may itself be a function of time).
\section{Entanglement of quasiparticles}\label{sec:entquasi}
Low energy excitations of generic interacting many-body systems are typically quasiparticle states,
e.g. fermionic quasiparticles of Landau-Fermi liquid, bosonic collective modes, including Goldstone modes (spin-waves, phonons) in broken symmetry states and various gapped excitations. Quantum information properties of these essentially single particle sectors can be understood in detail through explicit treatment of representantive wavefunctions.

In this section we show how the QI quantities introduced in the previous section behave for the very simple case of single-particle localized states in one dimension.
We compare it with a single-particle random state, and show how the localization properties of QI differ.
A generic state in the single-particle sector can be written as
\begin{equation}\label{single}
|\psi\rangle =\sum\limits_{r=1}^{n}\psi_{r}|r\rangle,
\end{equation}
where $|r\rangle$ denotes local excitation basis, e.g. in the case of ferromagnetic ground state of $n$ spins $S=1/2$
\begin{equation}
|r\rangle \equiv |\downarrow\rangle_{1}\otimes\cdots\otimes|\downarrow\rangle_{r-1}\otimes|\uparrow\rangle_{r}\otimes|\downarrow\rangle_{r+1}\otimes\cdots\otimes|\downarrow\rangle_{n}.
\end{equation}
Simple algebra shows that the entanglement entropy of any desired set of spins $A$ is given by
\begin{equation}\label{sm}
S_E (\rho_{A})= -p_{\textrm{A}}\log p_{\textrm{A}}-(1-p_{\textrm{A}})\log (1-p_{\textrm{A}})\;,
\end{equation}
where we have defined
\begin{equation} 
 p_{\textrm{A}}=\sum\limits_{r\in A}|\psi_{r}|^{2}\;.
\end{equation}
Formula (\ref{sm}) is a generic formula, no matter the state (\ref{single}) and subsystem $A$ we consider. The problem in the single excitation subspace always reduces to the computation of the reduced probability $ p_{\textrm{A}}$.

\subsection{Localized quantum states}
Localized wavefunctions are typically characterized by a  peak located at a random position $R_0$
and a localization length $\xi$, giving the rate of decay, e.g.
\begin{equation}
\psi_{r}\propto e^{- |r-R_0|/\xi}\;.
\label{eq:toypsi}
\end{equation}
It is useful therefore to compute and compare observables in a coordinate frame centered on  
\be
R_0\approx \langle r \rangle_\psi.
\ee 
While $\xi$ is strictly speaking defined by the decay of $\psi$ in the tail ($|r-R_0|\to \infty$) we may instead consider a ball of size $\xi_S$ centered on $R_0$ (i.e. $\beta(R_0,\xi_S)\equiv \{y:|y-R_0|<\xi_S \}$) whose entanglement entropy is maximal
\be
S(\beta(R_0,\xi_S))\equiv \log 2.
\ee
Although for simple featureless localized states (e.g. in Eq. \ref{eq:toypsi}) $\xi_S=2 \xi/\log 2$, this need not hold in more complicated cases where additional short range structures appear\footnote{For example, deep inside the Lifshits band tails of the continuum Anderson model. We thank S. Gopalakrishnan for suggesting this example to one of us.}. Quite generally, in the information theoretic sense, $\xi_S$ is a natural length scale to use to define the "localization volume".

\begin{figure}[h]
 \includegraphics[width=0.48\linewidth]{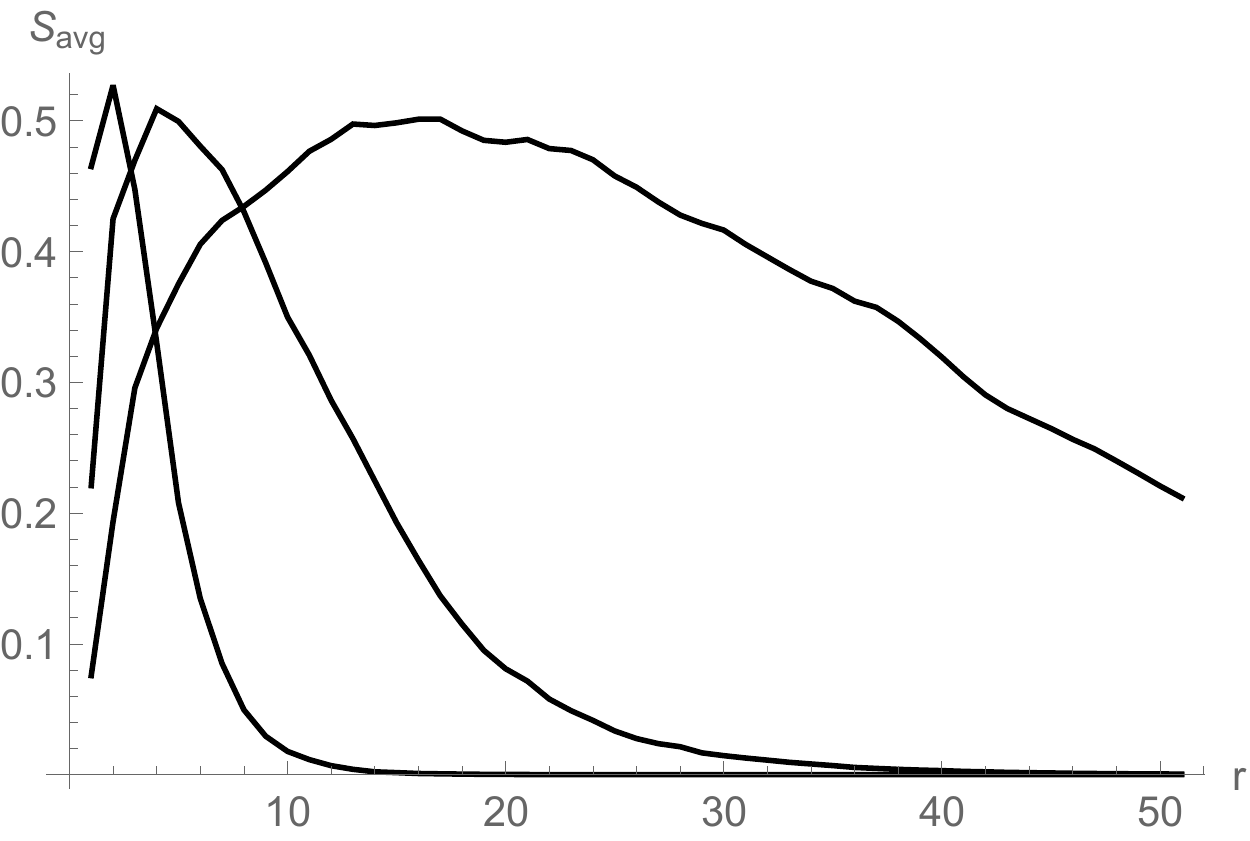}
 \includegraphics[width=0.48\linewidth]{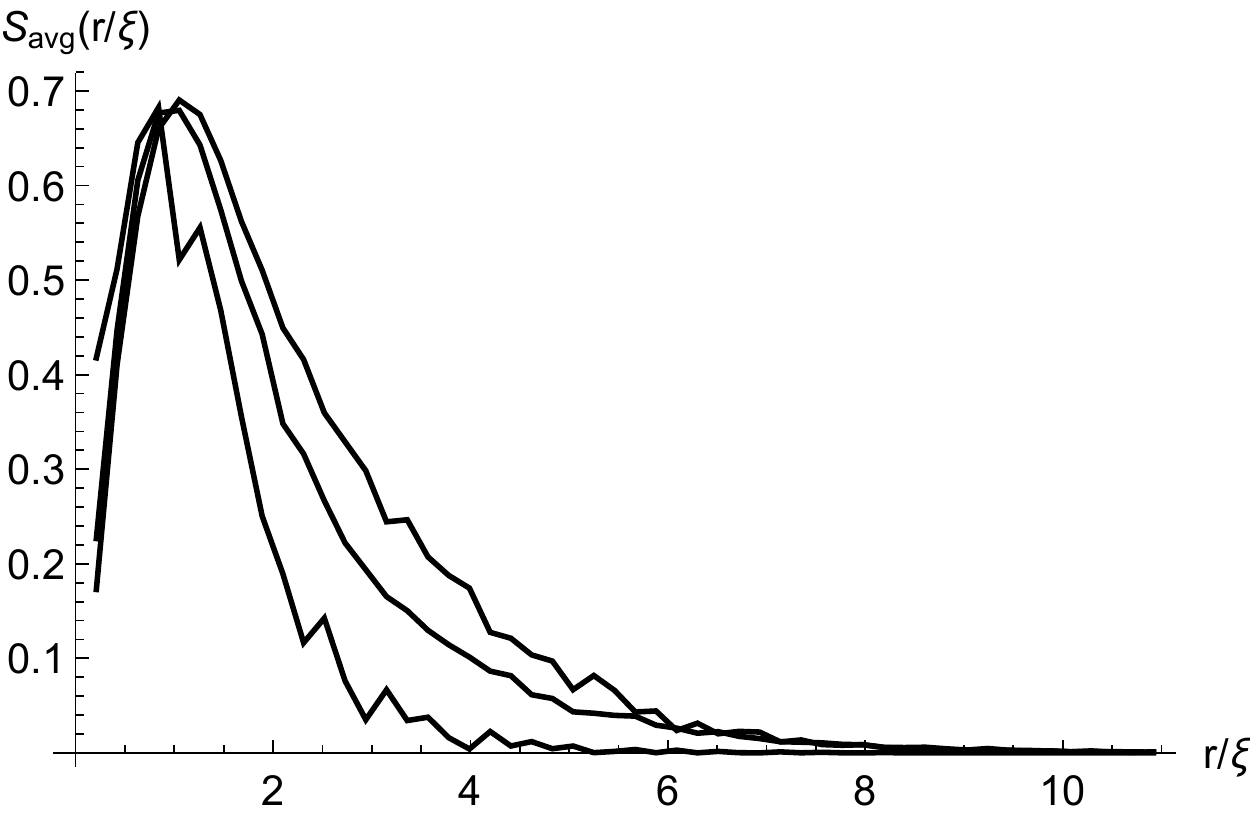}
   \caption{Average subsystem entropy for 3 disorder strengths $W=0.5, 1, 2$, averaged without (left) and with (right) rescaling with eigenstate specific $\xi_S$ extracted from the location of the peak for each eigenstate.}
    \label{fig:Swwoutrescaling}
\end{figure}

We illustrate the usefulness of $\xi_S$ using a generic lattice (Anderson) model with quenched disorder in site potentials, $V_j$ drawn from independent uncorrelated normal distribution of variance $W$
\be
H=\sum_{j=1}^n |j\rangle \langle j\pm 1| +V_j |j\rangle \langle j|.
\ee
Spectral (state-by-state) fluctuations in localization length necessarily broaden out the entropy peak in spectral averages (see left panel of Fig. \ref{fig:Swwoutrescaling}). 
Instead of simple averaging we can first obtain an estimate for $\xi_S$ for each eigenstate (from the location of its peak) and rescale the spatial coordinate w.r.t. to $\xi_S$ \emph{before} spectral averaging. As expected (see right panel of Fig. \ref{fig:Swwoutrescaling}) we restore the exact $\log 2$ peak. This rescaling also quantitatively "collapses" the typically non-universal but generic short distance structure of localized states.
As expected this rescaling cannot relate tails of the wavefunctions, which exhibit log-normal fluctuations and therefore should not be studied using simple spectral averages.

In discussing the codification volume $\beta_{\epsilon}$ we have a choice of subsystem $A$. For example, we may take it to be just the site at which the wavefunction peaks or the entire localization volume of radius $\xi_S$ about the peak.
In the scaling limit of large $L_\beta$ ($\epsilon\to 0$) we obtain simply from Eq. \ref{eq:ic}
\begin{align}
\label{eqn:scal}
&\epsilon
\simeq  \frac{2L_\beta}{\xi} e^{-L_\beta/\xi}\\
&L_\beta(\epsilon)\simeq \xi \log (\frac{2}{\epsilon} \log\frac{2}{\epsilon}\ldots),
\end{align}
for both choices of subsystem $A$.
As expected, this result is a scaling function of the dimensionless ratio of size of $\beta$ and the localization length.
Physically, $\epsilon$ may be thought of some externally imposed fixed amount of information loss (e.g. in experimental measurement or resolution context). We may then obtain the dependence of the codification volume on other properties of quantum states, e.g. $\xi$.

\subsection{Random delocalized single particle states}
At the other extreme
we can study the physical properties of a typical delocalized single particle state in the single particle by defining an effective random ensemble. Operationally, up to subleading corrections, 
the complex amplitudes associated with the state (Eq. \ref{single}) 
are gaussian random variables
\begin{equation}
[\psi_{i}\psi_{j}^{*}]=\Lambda \delta_{ij}\;,
\end{equation}
where $\Lambda$ is fixed by normalizing the state
\begin{equation}
[\langle\psi\vert\psi\rangle]=\sum_{i}[\psi_{i}\psi_{i}^{*}]=\Lambda n=1\;,
\end{equation} 
We can compute\cite{Magan2015} again the MI between the first site and subsystem $B$, composed by $m$ sites. For $m\gg 1$ we have:
\begin{equation}
I(1,B)\simeq \frac{\log N}{N}-\frac{1}{N}\log (\frac{1-\frac{m}{N}}{\frac{m}{N}})\;
\end{equation}
which shows no special spatial structure, as opposed to~(\ref{eqn:scal}). On the other end, for $m\sim\mathcal{O}(1)$, MI vanishes in the thermodynamic limit faster than entanglement entropy itself, a hallmarck of the delocalized phase. For example, for $m=1$ we have:
\begin{equation}\label{eqn:MIdel}
I(1,1)\simeq  \frac{2}{N}\log 2\;.
\end{equation}
The previous equations show that large scale entanglement structures certainly discern between localized and delocalized scenarios, signalling the phase transition together with the right scaling of the correlation length. But more interestingly, they also show a different scaling of the MI between adjacent points in the delocalized phase~(Eq. \ref{eqn:MIdel}) which vanishes in the thermodynamic limit, in contrast with the localized scenario. As stated in the introduction, this is an example of how properly analysed and sampled short range correlations are expected to be as sharp observables as long range entanglement structures. This is important since few site entanglement entropy and quantum information is measureable (inferable, to be precise) in principle in experiments, from complete tomography of the reduced density matrix of small subsystems.

\subsection{Dynamics}
While eigenstates provide a clear dichotomy between localized and extended behaviors, a more physically relevant setup for exploring transport (including transport of quantum information) is by preparing simple initial states with specific patterns of observables and allowing for them to evolve under action of the Hamiltonian. Such protocols may be thought of ``quantum quenches'' (switching between preparation and evolution Hamiltonians) and have been investigated recently using cold atomic gasses and theoretically in a  bewildering diversity of different physical situations. We expect the system to reach a steady state described by the so-called ``diagonal ensemble'', whose time-averaged density matrix is nearly diagonal in eigenbasis. This steady state need not possess true thermal correlations, in fact, examples of so-called GGE (generalized Gibbs ensemble) are well known, e.g. in integrable many-body systems. The character of the steady state density matrix is essentially determined by the initial amplitudes of eigenstates. For typically realized initial conditions of essentially classical product states, with completely localized pattern of excitations, these amplitudes are qualitatively different between localized vs. extended eigenbases. In the former case, the distribution of amplitudes is essentially saturated by a small, often vanishing, fraction of basis states, while extended eigenstates are sampled much more democratically. Thus, spectral localization-delocalization transition may be observed in the changes of steady state properties reflecting changes in ``participation fraction'' of eigenstates in the initial condition. Roughly speaking, on the delocalized side we expect time evolved protocol to provide access to eigenstate averages discussed above, making careful preparation of initial states unimportant (broadly consistent with Eigenstate Thermalization Hypothesis), while on the localized side time evolution alone is insufficient to correctly sample the eigenstates. Empirically, extensive sampling over initial conditions appears to mimic eigenstate averages although possibly not for all quantities.

Application of these ideas to single particle problems and also many-body problems with macroscopically inhomogeneous initial states are well known and conceptually straightforward (although perhaps not using multipoint entanglement measures defined above).  Importantly, these examples highlight natural connection between transport of physical observables (especially energy) and information, which is not surprising considering everyday experience (phone lines, radio) where elementary excitations' amplitude and/or phase are used to transport (classical) information, i.e. Anderson localization arrests \emph{all} forms of transport.

\section{spin chains}

We start by summarizing a few basic facts about quantum entanglement of interacting systems. First and foremost, strong volume-like entanglement of eigenstates is the microscopic underpinning of the so called Eigenstate Thermalization Hypothesis, a modern quantum reincarnation of the conventional Ergodicity Hypothesis that enables passing from classical many-body dynamics to statistical thermodynamics descritpion. Put simply, in such ergodic regime quantum (either von Neumann or Renyi) entropy of a subsystem of an isolated system converges to the statistical (microcanonical) value of entropy.  Starting from weakly entangled initial states intrinsic many-body dynamics generates entanglement at a constant rate (so called generalized Lieb-Robinson velocity) until volume law saturation limit is reached.  This much (and slightly more) is known about entanglement in the delocalized regime of the phase diagram.
While it was appreciated early on that onset of localization should thwart strong volume-like entanglement (viz-a-viz breakdown of thermalization), e.g. most likely by producing area laws even in excited eigenstates, it came as a bit of a surprise that simple ``classical'' product states generalically evolve into volume-entangled linear combinations of weakly entangled eigenstates (with surface-like entanglement). This phenomenon is most simply understood as many-body dephasing of localized and ``conserved'' excitations by residual interactions. Temporal growth of entanglement in this regime is parametrically slower -- logarithmic in time for generic short-ranged interactions (or powerlaw for long-range interacting models).  This in turn can lead to non-universal powerlaw relaxation of certain observables, also interaction dependent powerlaws in spectral functions, distinct from conventional Mott-AC-type spectra known from studies of Anderson localized states.

In what follows we explore these basic patterns of entanglement using the multipoint measures defined above, complementing several recent works \cite{Campbell2016,Detommasi2016,Goold2016} from various perspectives. The discussion will be divided into three parts: (i) Focussing on two adjacent sites we will use their mutual information to study heterogeneity of entanglement in disordered eigenstates. We address the question of what sorts of many-body states may be optimal for quantum information processing; (ii) We then contrast the dependence of mutual information between a single site and progressively larger subsystems in localized vs. diffusive vs. purely random (in some sense maximally extended)  states, observing clear differences among all three types. These results \emph{quantify} the notion of an intrinsic ``bath'' a given spin loses information to; (iii) Finally, we explore structure of stationary states which, in many-body localized regime, exhibit considerable history dependence w.r.t. to initial conditions and can exhibit delocalization of quantum information correlating with partial relaxation of few body observables.

\subsection{Entanglement vs. quantum information}
Considerable amount of work has been exerted already to link many-body localization and scaling of bipartite entanglement. Here we pose and explore a different question -- is entanglement ``good'' for information transfer? What is clear, upon some reflection, is that very strong volume-like entanglement cannot possibly be fruitfully associated with quantum information. One way to see this is by appealing to ETH to estimate MI between two nearby (even nearest neighbor) sites -- the answer is zero up to finite size effects \cite{uscod}, anagolously see Eq.~(\ref{eqn:MIdel}) derived in the previous Section. Alternately, quantum spin-chains may be thought of literally as information channels and presence of strong entanglement is synonymous with \emph{generation} of noise and vanishing channel capacity\footnote{A. Scardicchio, V. Oganesyan, unpublished}. It is also rather clear that very strong disorder will quench any entanglement and suppress MI and channel capacity as well. Therefore we are naturally led to the proposal that optimal states for quantum information transport reside at intermediate disorder strength, possibly in the transition regime.
\begin{figure}[h]
  \includegraphics[width= \linewidth]{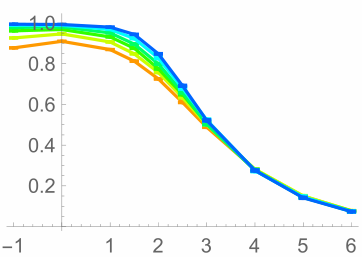}\llap{\raisebox{3cm}{\includegraphics[width= 0.4\linewidth]{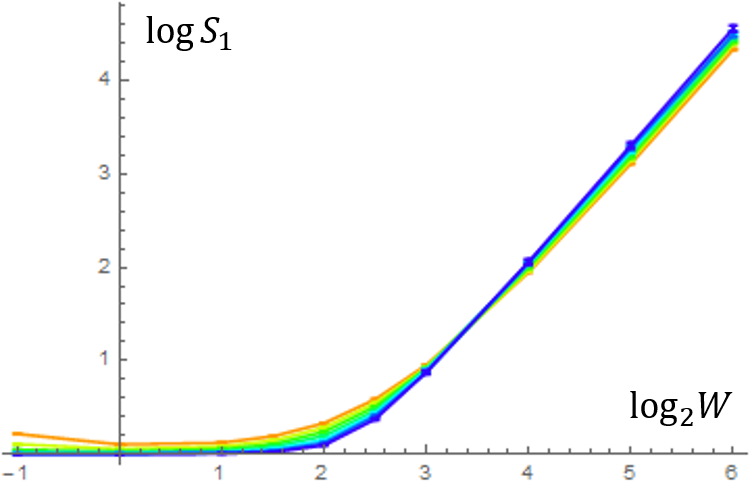}}}
  \includegraphics[width= \linewidth]{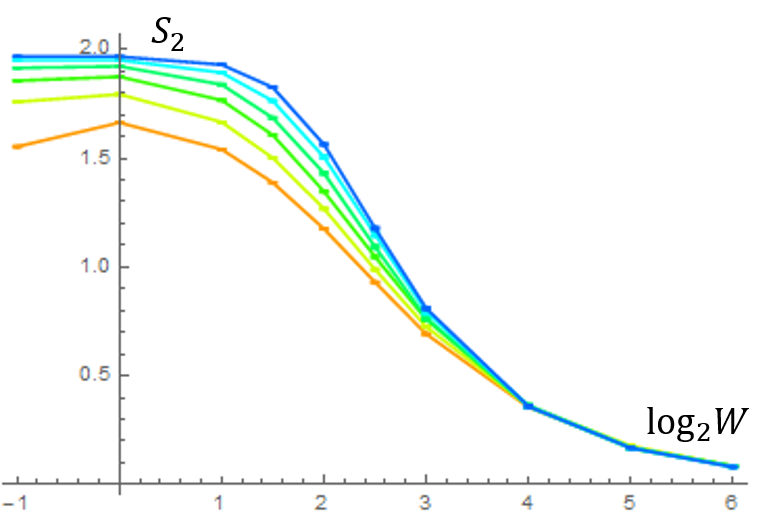}
  \includegraphics[width= \linewidth]{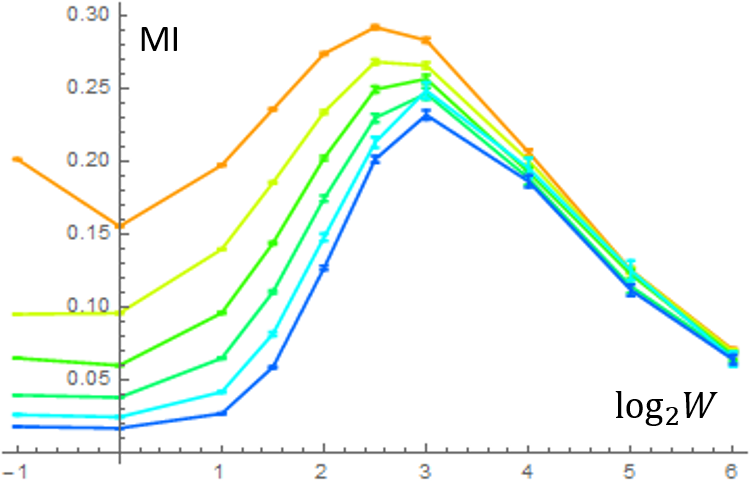}
   \caption{(a) Single site entanglement entropy (normalized by log 2) computed for Hamiltonian in Eq. \ref{eq:HamXY}, inset -- log-log plot to illustrate simple perturbative scaling of entanglement in the localized phase; (b) two-site entanglement entropy (normalized by log 2); (c) Mutual information.}
\label{fig:s1}
\end{figure}
To see this transition we find the optimal disorder strength for communication protocols by computing the MI between nearby points.  The connection of the short range MI to fidelity rates for quantum information transport \cite{Lorenzo2015,Apollaro2015}  is left for future work as well as the possible information transfer to a coupled degree of freedom \cite{giorgi2010,paganelciuk3}.In Fig~(\ref{fig:s1}) we show the one site entanglement, two sites entanglement, and MI between two nearby sites in a spin chain whose Hamiltonian is 
\be
H=\sum_{j=1..L-1} \sigma^\pm_j \sigma^\mp_{j+1}+\bf{h}_j\cdot\bf{\sigma}_j,
\label{eq:HamXY}
\ee where $\sigma_j$'s are Pauli matrices and $\bf{h}_j$ only has $x$ and $z$ components, each drawn randomly from a normal distribution of width $W$.
As we increase the disorder strength we lose ergodicity, and entanglement entropy decreases -- this is expected and clear from first two panels in Fig. \ref{fig:s1}.  Finite size effects preclude addressing whether scaling behavior is observed near the transition, e.g. whether the entropy goes to the thermal value smoothly or discontinuously. 
Mutual information shows a more interesting behavior, with a peak at a critical disorder strength (which for this model we corroborated using level-statistics\footnote{E. Kapit and V. Oganesyan, in preparation}). This feature was in fact anticipated by the heuristic discussion above but is not immediately apparent by looking at entanglement entropies alone. We propose that this peak and its associated critical disorder strength are connected with the physics of communication protocols in MBL phases, a direction we plan to explore in the future. From a different perspective it is interesting to explore the connection between this approach to the MBL phase transition and the analysis of the entanglement entropy at the critical point and through the phase transition performed in Ref. \onlinecite{Khemani2016}.

\subsection{Intrinsic bath size vs. disorder}
For the remainder of this Section we turn to the commonly studied random field Heisenberg chain\cite{znidaric2008,Pal2010a}
\begin{equation}\label{Hamil}
H= \sum_{i=1}^{n-1} J_{\perp}(\sigma_{i}^{x}\sigma_{i+1}^{x}+\sigma_{i}^{y}\sigma_{i+1}^{y})+J_{z} \sigma_{i}^{z}\sigma_{i+1}^{z}+\sum_{i=1}^{n} h_{i}\sigma_{i}^{z}\;,
\end{equation}
where $J_{z}/J_{\perp}=0.2$ and $h_{i}$ is a random variable with uniform probability distribution in the range $[-\eta,\eta]$. In the numerical simulations we diagonalize the Hamiltonian exactly, with $n=8$ spins. In all cases we average over $1000$ repetitions.  When studying the deep MBL phase \cite{huse} we will set $\eta=6$.

\begin{figure}[h]
  \includegraphics[width= 5.5cm,angle=-90]{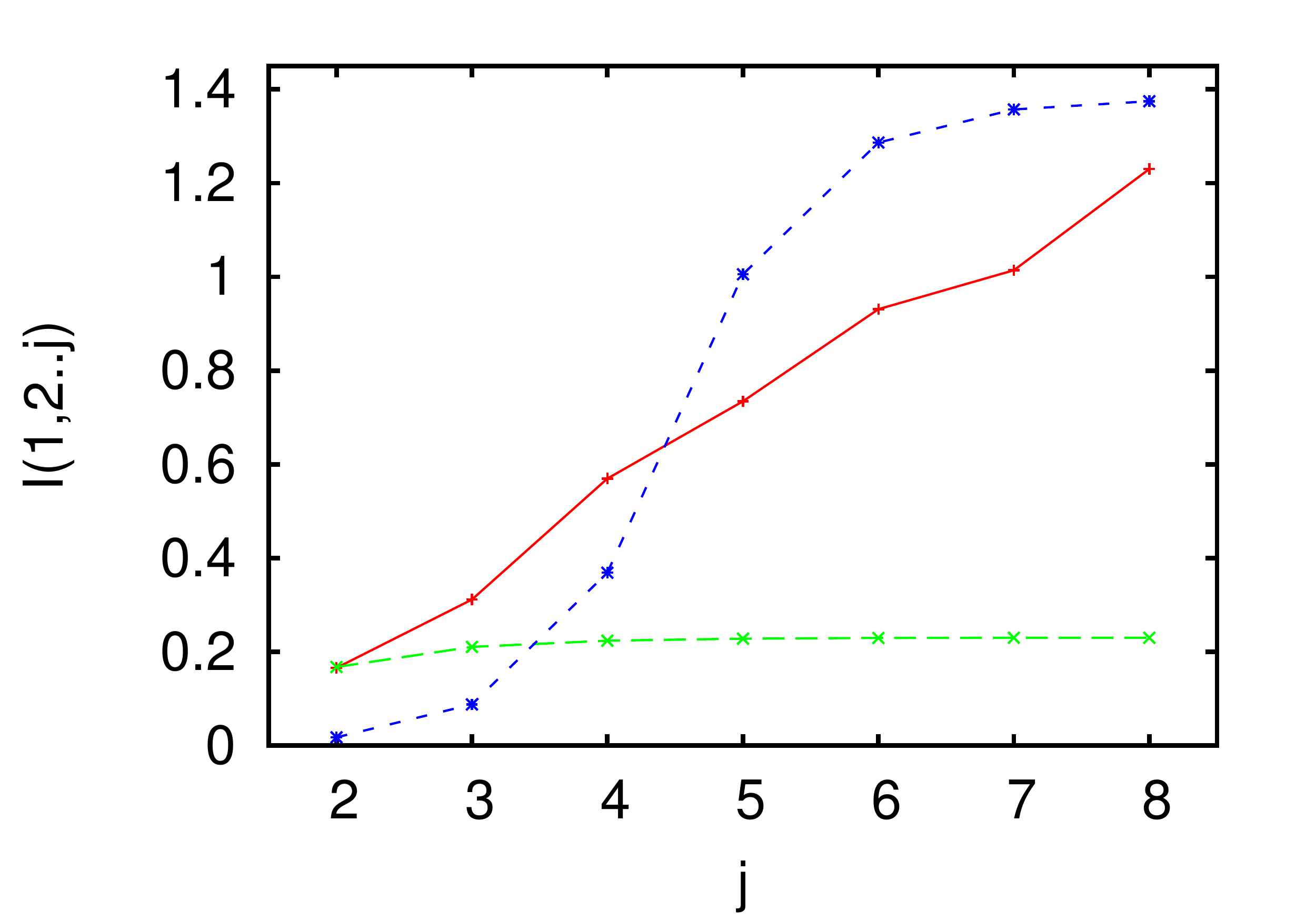}\\
   \caption{The structure of Mutual Information between the first spin and different size adjiacent blocks of spins associated to:
    the average over the eigenstates of the MBL Hamiltonian (\ref{Hamil}) (green); a random state in the Hilbert space (blue);
   the average over the eigenstates of a non-integrable Hamiltonian (red) .
   \\
 }
    \label{eigstates}
\end{figure}

We examine the structure of the "peak-to-volume" MI  (PVMI) $I(1,B)$ between a point at the end of the chain and adjacent group of sites B by averaging over all eigenstates of each sample Hamiltonian, and then by averaging over several disorder configurations. The result (deep inside MBL phase) is depicted in Fig.  \ref{eigstates} (green line). Typically, the small amount of information shared by spin $1$ is mostly contained in spin $2$, and adding more spins increase very little the MI
\begin{equation}
I(1,\bar{1})=2S_{1}\simeq I(\textrm{1},\textrm{2}).
\end{equation}
The QI associated to each degree of freedom is  \emph{localized} in this precise way in disordered eigenstates, the deficit information decreasing very fast as we increase $B$, probably with a law of the type~(\ref{eqn:expodecay}), which we cannot resolve due to number of spins considered.

Naively, one would think this is just another way to look at the entanglement area laws found for disordered eigesntates\cite{nayak}. The advantage of this approach is that the deficit information is directly connected with the correlation length of the system, the main challenge here being the numerical study of Hamiltonians with a larger number of spins, leading to a law of the type~(\ref{eqn:expodecay}) which would provide the scaling law of the correlation length, an approach which was explored recently in \cite{Detommasi2016}.

The average structure of the MBL phase is markedly different from the structure of a thermal/random state, computed in \cite{uscod} and shown in Fig  \ref{eigstates} with color blue, or with the average structure between the eigenstates of the local non-integrable Hamiltonian studied in Ref. \onlinecite{uscod}, shown in the figure in red. In ergodic phases, quantum information is strongly delocalized over the system. This is seen by the increase of $I(1,B)$ as we increase $B$.

\subsection{On initial state dependence of information delocalization}
In this last part we study aspects of multipoint entanglement in non-equilibrium unitary processes. To this end we consider time-evolved product states drawn from continuously deformed ensembles of initial conditions realizeable experimentally \cite{huse}. We will be interested in the (de)localization properties of QI at stationarity.

The family of initial states are distinguished by the expectation value of the $z$ component of the single site spin. Defining:
\begin{equation}
|\psi\rangle_{\theta}=\cos(\frac{\theta}{2})|\downarrow\rangle+i\,\sin(\frac{\theta}{2})|\uparrow\rangle\;,
\end{equation}
we study the following family of initial states parametrized by $\theta$:
\begin{eqnarray}\label{family}
&|\Psi_{\theta}\rangle=|\psi\rangle_{\theta}\otimes|\psi\rangle_{\pi-\theta}\otimes|\psi\rangle_{\theta}\otimes|\psi\rangle_{\pi-\theta}\otimes & \nonumber \\ &|\psi\rangle_{\theta}\otimes|\psi\rangle_{\pi-\theta}\otimes|\psi\rangle_{\theta}\otimes|\psi\rangle_{\pi-\theta}&\;,
\end{eqnarray}
with $0\leq \theta \leq \frac{\pi}{2}$. The interesting aspect of this family is that it smoothly interpolates between the ``antiferromagnetic'' state, for $\theta=0$,
and a product of all spins pointing in the positive $y$ direction for $\theta =\frac{\pi}{2}$. For these two extreme cases, the evolution of the multipartite entanglement structure  was studied in \cite{uscod} for the case of a quantum chaotic Hamiltonian. In such a case, the multipartite entanglement structure of both states evolves towards the structure of a random state, blue line in Fig~(\ref{eigstates}).

\begin{figure}[h]
\includegraphics[width= 5.5cm,angle=-90]{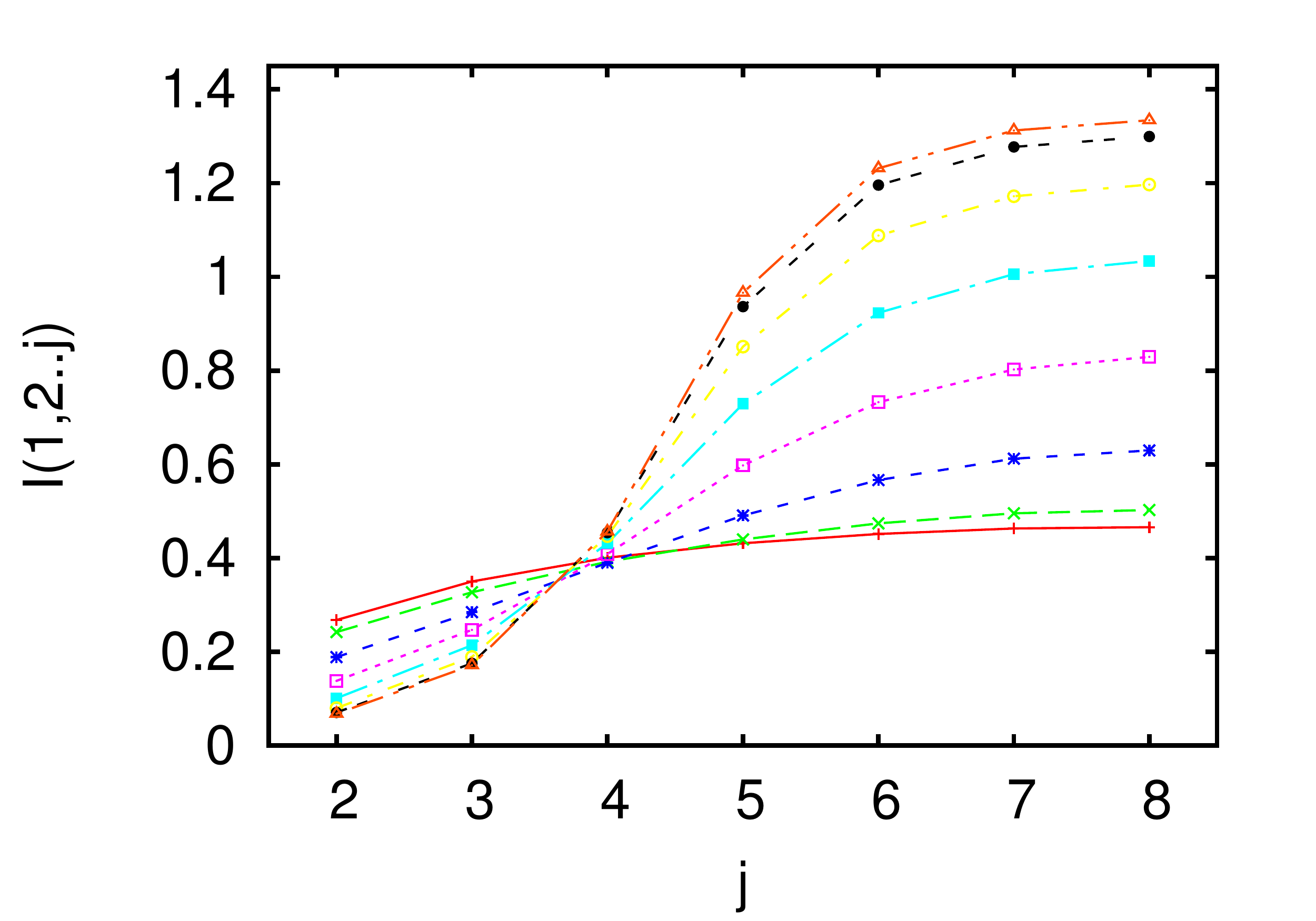}\\
\caption{Peak-to-volume MI at stationarity. Each curve corresponds to one of the initial states parametrized in (\ref{family}), with $\theta_{k} =\frac{k \pi}{14}$ with $k=0,\ldots, 7$. The plot of the MI structure of a random state (red color) is shown for comparison. All curves collapse to the random expectation in the delocalized phase.}
\label{mutuallocalized}
\end{figure}

The peak-to-volume MI at stationarity for each initial state $|\Psi_{\theta} \rangle $ after the evolution driven by the Hamiltonian (\ref{Hamil}) with disorder $\eta=6$, for $200$ repetitions is shown in Fig~(\ref{mutuallocalized}). The antiferromagnetic $\theta =0$ case is the lowest line. As we increase $\theta$ the structure of MI keeps approaching the structure of the random state \cite{uscod}, which appears also in the figure as the highest line. In a chaotic/delocalized phase, all initial states would collapse to the random expectation.

\begin{figure}[h]
  \includegraphics[width= 5.5cm,angle=-90]{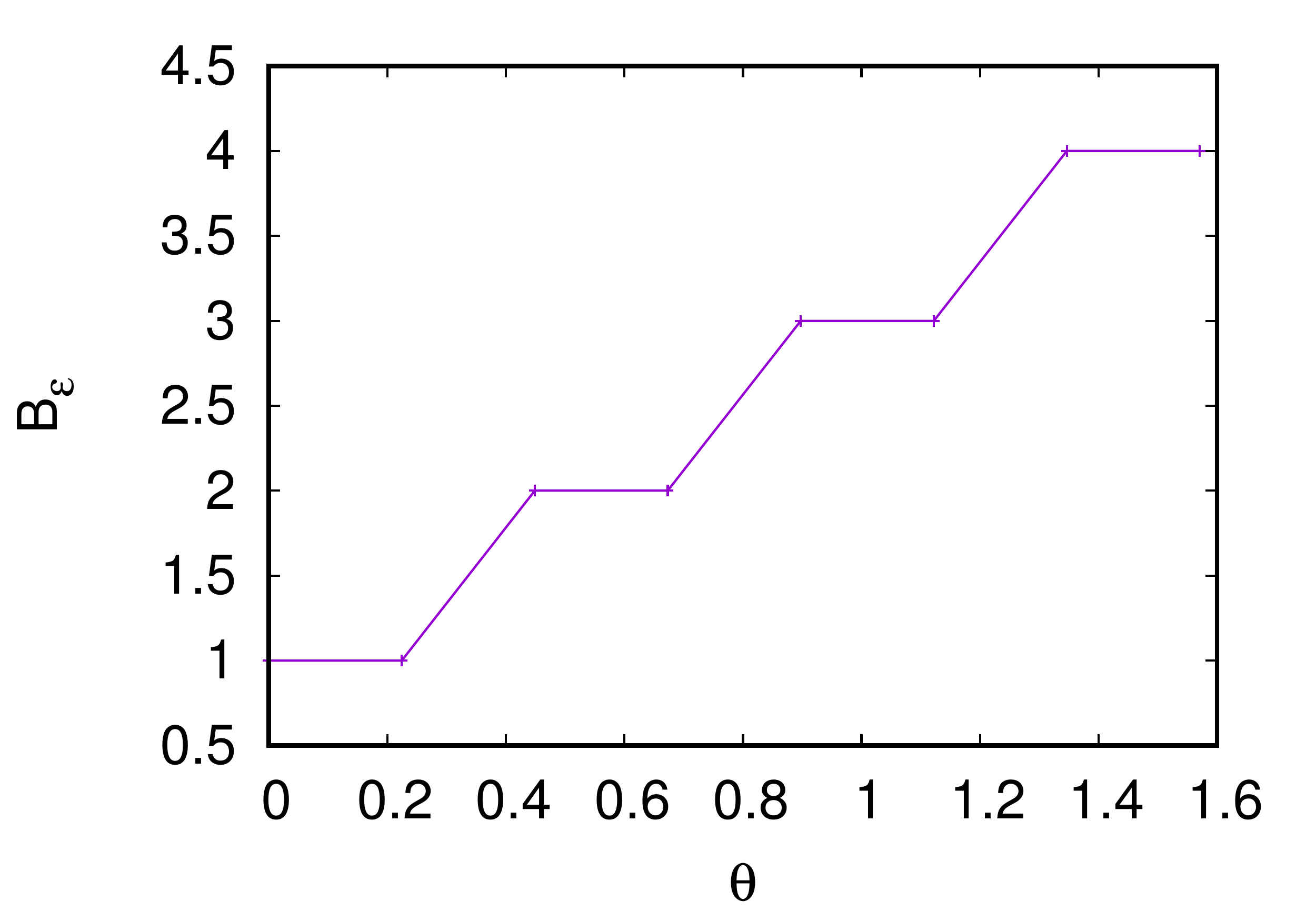}\\
    \caption{Codification Volume, for $\epsilon=0.3$, as a function of $\theta$ at stationarity for a many-body-localized Hamiltonian.}
    \label{cvangle}
\end{figure}

From~(\ref{mutuallocalized}) we conclude that the amount of delocalization of quantum information in the MBL phase is directly controlled by $\theta$. The delocalization properties can be also studied by stablishing certain tolearnce $\epsilon$ and finding the codification volume $B_{\epsilon}$. For $\epsilon=0.3$ the result is given in  Fig. \ref{cvangle}. 

This non-trivial and somewhat unexpected dependence on the initial state should have implications for quantum comunications protocols as well. As described in the previous section, too much delocalization of quantum information due to big amounts of entanglement is bad for comunication protocols, since this is equivalent to big amounts of noise. On the other hand, too much information localization avoids information transmission through the quantum channel. We are led to conclude that intermediate values of $\theta$ are better suited for these comunication protocols to work within the MBL phase. An important open question is to find out if there is a sharp value of $\theta$ dividing the family of initial states (\ref{family}) into localizable/delocalizable families in the thermodynamic limit. This value could potentially be measured experimentally.

Another interesting unexpected aspect of Fig~(\ref{mutuallocalized}) is the crossing point at $j\simeq 4$. The behavior of the crossing point as a function of $\eta $ might sheld novel light into the scaling properties of the MBL phase transition. We leave this interesting topic for future research.

At any rate, Fig~(\ref{mutuallocalized}) shows that quantum thermalization might occur in systems that do not satisfy ETH, as the one we are considering, and might be strongly dependent on the initial state considered and not only in the entanglement properties of energy eigenstates.

\subsubsection{Connecting information delocalization with experiments}

\begin{figure}[h]
  \includegraphics[width= 5.5cm,angle=-90]{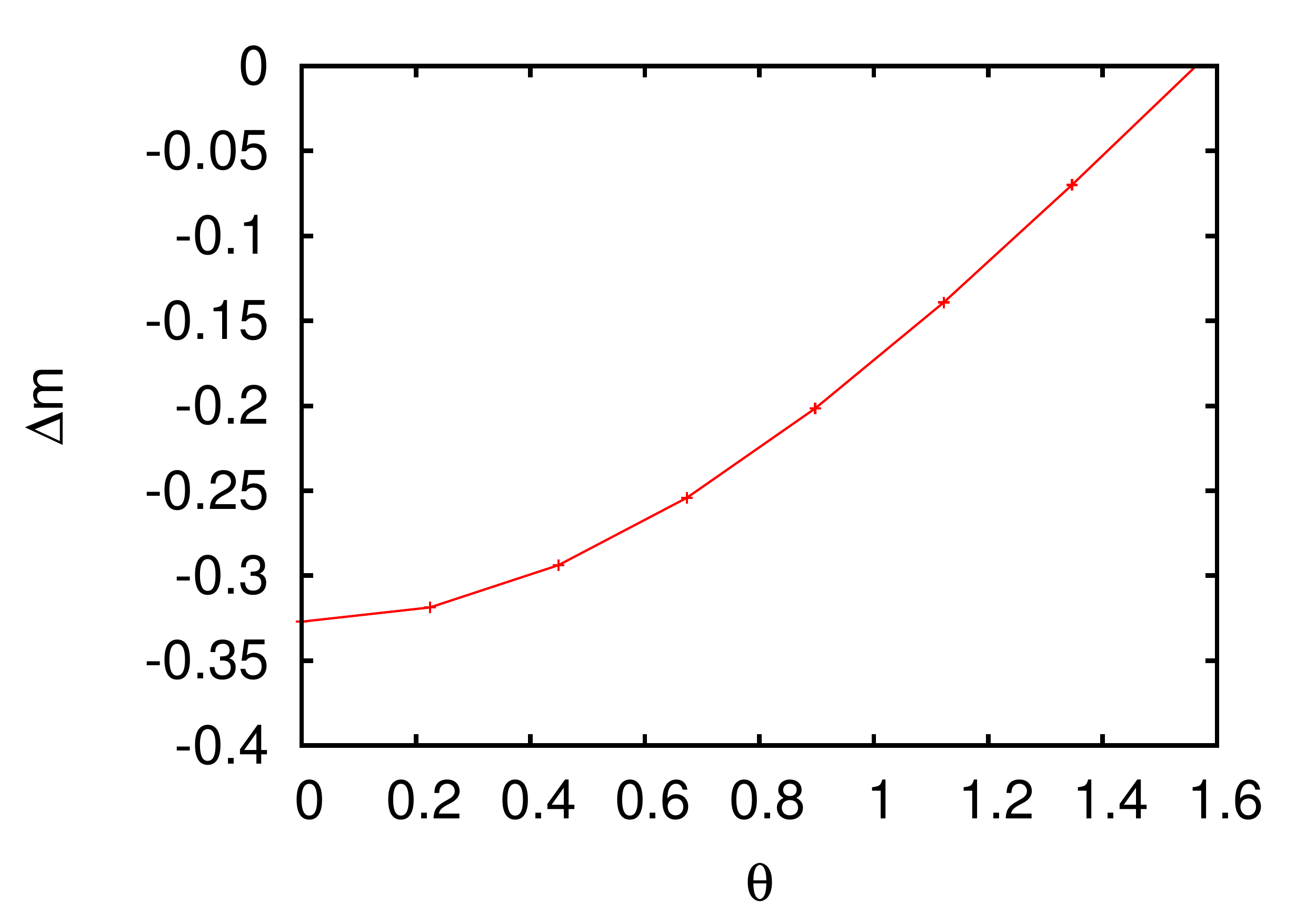}\\
    \caption{Magnetization imbalance as a function of $\theta$ at stationarity.}
    \label{Imbangle}
\end{figure}

Recently \cite{schreiber2015}, MBL has been observed 
in a fermionic system prepared in an optical lattice in a charge density wave configuration. 
To characterize the ergodicity 
breakdown,  the \emph{charge imbalance} was studied. For our spin system, the analogous quantity is the magnetization imbalance, defined as
\begin{equation}\label{imbalance}
\Delta m= \sum_j \langle \sigma^z_{2j} -\sigma^z_{2j+1}\rangle .
\end{equation}
This quantity clearly vanishes in a thermalization process. Using the same Hamiltonian and family of initial states, as in the previous section, we compute the steady state magnetic imbalance~(\ref{imbalance}) in Fig. \ref{Imbangle}.  The behavior paralells that of the  PVMI and codification volume $B_{\epsilon}$ studied in the previous section.  The magnetic imbalance vanishes as the initial state approaches $\theta=\frac{\pi}{2}$, i.e when the MI approaches the ergodic random state behavior, or when the $B_{\epsilon}$ is maximal.

A natural question arises: why to compute these complicated information quantities after all? To answer this question let us notice that although suitable for these non-equilibrium scenarios, the magnetic imbalance is not able to pin-point the MBL phase transition in the spectrum of eigenstates, since the average over the spectrum would force the imbalance to vanish, as in an ergodic or quantum chaotic Hamiltonian. On the other hand, as we have seen above, the MI and $B_{\epsilon}$ provide a universal framework, valid for the spectrum of eigenstates and non-equilibrium unitary processes, to study the MBL phase transition and the scaling of the correlation length deep inside the MBL phase.

\section{Summary and future directions}

In this article we have used several spatial resolved QI quantities to study the break-down of ergodicity in interacting disordered systems and also to characterize the way quantum information is able to spread in the disordered phase.
To this aim we have focused on MI, specifically on the MI between nearby points, PVMI and the CV, well defined notions which able to describe the spreading of the QI in a many-body system.

We have  first analyzed the behavior of the QI quantities  of interest for a 1D  spin-$1/2$  system, where a single quasiparticle is excited. Two extreme cases has been considered: the  \emph{localized}  and the  \emph{typical}  (random) state. The former emerges when the ergodicity is broken while the latter represents the results of an ideal thermalization  process.  The results of section~(\ref{sec:entquasi}) show that in the single-particle scenario, QI quantities provide a benchmark for both localization and thermalization phases, including the indentification of the correlation length.

We then extended our analysis to the many-particle scenario, the idea being to characterize the MBL by the QI quantities as in the single-particle case. We first argued that, given the information localization in the MBL phase together with the expected vanishing of channel capacity in an ETH phase (due to the strong noise produced by volume law entanglement), the optimal regime for quantum comunication protocols lies at intermediate values of the disorder strength. The approximate value of this optimal disorder can be found by computing the MI between neighbouring sites, a quantity that shows a peak across the delocalized vs localized phase transition. It is left for future work to compare these results with more rigorous approaches to state transport in spin chains, such as those performed in \cite{Apollaro2015,Lorenzo2015} using fidelity measures.

We then studied the PVMI  for the eigenstates of the theory, extracting the typical structure by averaging over all of them. The results are markedly different from the thermal/random case, or the average over the eigenstates of a non-integrable Hamiltonian. These results confirm from this novel perspective the results found in  \cite{nayak}. In the MBL phase ETH is broken, and information is localized within nearest neighbors in the Hamiltonian eigenstates. An advantage of studying PVMI and the deficit information is that the correlation length of the system can be extracted more easily, by fitting the numerics with effective laws of the type~(\ref{eqn:expodecay}), if computations with a higher number of spins can be accomplished.

Finally, in order to study the possible ergodicity breaking, the information propagation during non equilibrium unitary processes has been also considered. We have studied the stationary regimes corresponding to the late time unitary evolution of a family of initial states. The results seem to point out that localization of QI depends strongly on the initial quantum state. Indeed, for some initial states QI clearly propagates along the chain, the structure of MI being very similar to that of a random state. Since the system violates ETH, this result may leave ETH as a sufficient condition for quantum thermalization, but not a necessary one. To connect the QI quantities with directly measurable observables, and in analogy with the recent experimental work  \cite{schreiber2015}, we studied the magnetization imbalance in the stationary state for different $\theta$s, finding an analogous initial state dependence.

We are not aware of any unified framework, on par with ordinary hydrodynamics or quantum field theory, capable of capturing spatial and temporal structures of entanglement and quantum information discussed in this manuscript and other related recent efforts\cite{Campbell2016,Detommasi2016,Goold2016}.  In many-body systems with disorder in particular there is a preferred basis (i.e. real space) in which information may be localized possibly allowing for an effective classical statistical description, possibly akin to roughening or jamming with thermal phas. Whether or not these analogies can help elucidating, e.g. critical properties


\section{Acknowledgments}

We acknowledge partial support from MCTI and UFRN/MEC (Brazil). Research conducted with the aid of a high performance computer system, 
of the International Institute of Physics - UFRN, Natal, Brazil. J.~M is supported by the Delta-Institute for Theoretical Physics (D-ITP) that is funded by the Dutch Ministry of Education, Culture and Science (OCW). SP is supported by a Rita Levi-Montalcini fellowship of MIUR. 

\bibliographystyle{apsrev4-1} 
\bibliography{bibusmanycodif,C:/Users/vadim/Dropbox/library}

\end{document}